\documentclass[prd,twocolumn]{revtex4}
\usepackage{bm}
\usepackage{epsfig}
\usepackage{amsmath}
\newcommand{\be}{\begin{equation}}
\newcommand{\ee}{\end{equation}}
\newcommand{\ba}{\begin{eqnarray}}
\newcommand{\ea}{\end{eqnarray}}
\newcommand{\bes}{\begin{subequations}}
\newcommand{\ees}{\end{subequations}}

\newcommand{\bi}{\begin{itemize}}
\newcommand{\ei}{\end{itemize}}
\newcommand{\gev}{~{\rm GeV}}

\begin{document}
\title{Experimental constraints on fourth generation quark masses}
\author{P.Q. Hung}
\email[]{pqh@virginia.edu}
\affiliation{Dept. of Physics, University of Virginia, \\
382 McCormick Road, P. O. Box 400714, Charlottesville, Virginia 22904-4714, 
USA}
\author{Marc Sher}
\email[]{mtsher@wm.edu}
\affiliation{Dept. of Physics, College of William and Mary,\\  Williamsburg,
Virginia 23187, USA}
\date{\today}
\begin{abstract}
The existing bounds from CDF on the masses
of the fourth generation quarks, $t^\prime$ and $b^\prime$, are 
reexamined.  The bound of $256$ GeV on the $t^{\prime}$ mass assumes 
that the primary decay of the $t^{\prime}$ is into $q+W$, which is 
not the case for a substantial region of parameter space.  The bound 
of $268$ GeV on the $b^{\prime}$ mass assumes that the branching ratio 
for $b^{\prime}\rightarrow b+Z$ is very large, which is not only not 
true for much of parameter space, but is {\em never} true for 
$b^{\prime}$ masses above $255$ \gev.   In addition, it is assumed that 
the heavy quarks decay within the silicon vertex detector, and for 
small mixing angles this will not be the case.  The experimental bounds,
including all of these effects, are found as a function of the other 
heavy quark mass and the mixing angle.

\end{abstract}
\pacs{}
\maketitle

The question of whether or not there exist quarks and leptons
beyond the known three generations has generated
a number of theoretical and experimental investigations\cite{fhs,later}. 
Although direct
experimental constraints
did not (and do not) rule out a heavy fourth generation, until recently
electroweak precision data appeared to disfavour its 
existence. In addition, there is a strong prejudice against
quarks and leptons beyond the third generation which is
usually paraphrased by the question: Why is the
fourth neutrino so much heavier ($m_{\nu_4} > M_Z/2$)
than the other three? Of course, one
can very well imagine several scenarios in which this
can ``easily'' be accomplished since much is yet
to be learned about neutrino masses. In the end, it will
be the verdict of experiments which will be
the determining factor. 

There is still
the question: Why should one bother with a fourth
generation? There might be several answers to that
question. First, it is possible that a heavy fourth 
generation might help in bringing the $SU(3) \otimes
SU(2)_L \otimes U(1)_Y$ couplings close to a unification point
at a scale $\sim 10^{16}$ in the simplest non-supersymmetric
Grand Unification model $SU(5)$ \cite{hung}. Second,
its existence might have some interesting connections
with the mass of the SM Higgs boson \cite{kribs}.
Last but not least, there is no theoretical reason
that dictates the number of families to be three. We
still have no satisfactory explanation for the mystery
of family replication.

Recent reexaminations \cite{kribs,he} of electroweak precision data
led to the conclusion that the possible existence
of a fourth generation is not only allowed but its mass range is
correlated in an interesting way with that of the Higgs boson
in the minimal Standard Model (SM). In \cite{kribs},
the masses of the fourth generation quarks ($t^\prime$ and $b^\prime$) are taken to
be bounded from below at around $258$ GeV. This lower bound
was taken from CDF, who bounded the $t^{\prime}$ mass\cite{cdft} and 
the $b^{\prime}$ mass\cite{cdfb}.  However, CDF made a number of
assumptions. For instance,
it was assumed that $B(b^\prime \rightarrow b + Z) = 100\%$ which
yields a lower bound $m_{b^\prime} > 268$ GeV.   As we will show
below, this assumption is not only unjustified, but is in fact false 
for $m_{b^{\prime}} > 255$ GeV.   As a result,
the CDF bound weakens considerably\cite{caveat}.  Furthermore, we
show that there exist unexplored regions in the mass-
mixing angle parameter space which could prove very
useful to future searches.

Most of the present mass bounds on the fourth generation
quarks involved searches done within about $1$ cm from
the beam pipe, inside the silicon vertex detector of
CDF. The customary focus is on the decay of
$t^\prime$ and $b^\prime$ into quarks of the first three generations,
in particular into the $b$ quark. As noted above, for example, the
most recent bound on the $b^\prime$ mass by CDF was obtained
by searching for the decay mode $b^\prime \rightarrow b + Z$.
Similarly, the $t^\prime$ search focused on the decay mode
$t^\prime \rightarrow q+W$.

Let us first look at the $t^\prime$ decay.  One can consider
two regions: (I) $m_{t^\prime} \leq m_{b^\prime}$, and (II)
$m_{t^\prime} > m_{b^\prime}$. For (I), it is obvious that
the main decay mode will be $t^\prime \rightarrow q+W$.  For (II), whether 
or not
$t^\prime \rightarrow q+W$ dominates over $t^\prime \rightarrow b^{\prime}+W$ 
depends on the $b^\prime$ mass and the mixing angle
$\theta_{bt^{\prime}}$.   In this region, we calculate the branching 
ratio for $t^\prime \rightarrow q+W$ as a function of the parameters.  
If it is lower than $60\, \%$, the CDF bound will not apply (the 
choice of $60\, \%$ is based on viewing the
graph in Ref.\cite{cdft}, and is somewhat arbitrary without a more detailed 
analysis--since the region shown is on a logarithmic scale, changing 
that number will not noticeably affect the result).   If the 
$b^{\prime}$ mass is between $m_{t^{\prime}}-M_{W}$ and 
$m_{t^{\prime}}$, then the three body decay predominates;
whereas if it is lower than $m_{t^{\prime}}-M_{W}$, the two-body 
decay will dominate.   The decay rate is given in Refs. 
\cite{hou,bigi} and repeated (in this specific context) in equation [52] of Ref. 
\cite{fhs}.

In addition, even if the $t^\prime \rightarrow q+W$ decay does 
dominate, CDF assumed that it decayed within approximately 1 
centimeter from the beam pipe.  For very small mixing angles, this 
will not be the case.   Of course, for extremely small mixing angles, 
such that the decay length is greater than about 3 meters,
the $t^{\prime}$ will appear to be a stable particle and can be ruled 
out (above some mass) by stable quark searches.  

All of these results are plotted in Fig. 1, 
\begin{figure}
\includegraphics[clip,angle=0,width=9cm]{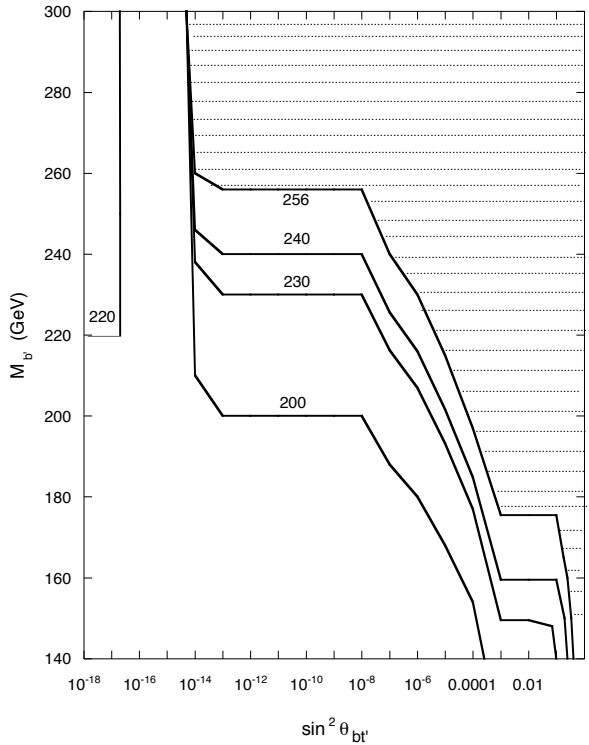}
\caption{Bound on the $t^{\prime}$ mass in the 
$m_{b^\prime}-\sin^{2}\theta_{bt^{\prime}}$ plane. The shaded
region corresponds to the CDF lower bound of $256$ GeV.}
\end{figure}
where we plot the bound on the $t^{\prime}$ mass in the 
$m_{b^\prime}-\sin^{2}\theta_{bt^{\prime}}$ plane.
There are three distinct regions. The shaded region above
and to the right of the curve with $m_{t^\prime}= 256$ GeV  
represents
the CDF lower bound on $m_{t^\prime}$. In this region, the CDF bound 
applies. Below the shaded region, the curves correspond to the new 
lower 
bound on $m_{t^{\prime}}$ from CDF based on the requirement that the 
$t^\prime \rightarrow q+W$ decay is dominant.
These curves all end to the left at
$\sin^{2}\theta_{bt^{\prime}} \sim 6\times 10^{-15}$. This
corresponds to a decay length of approximately 1 cm. (Let us
recall that present searches are sensitive to decays which
occur very close to the beam pipe to a distance of about 1 cm.)
To the left of those curves lies an {\em unexplored
window} situated between $\sin^{2}\theta_{bt^{\prime}} \sim 6\times 
10^{-15}$
and $\sin^{2}\theta_{bt^{\prime}} \sim 2\times 10^{-17}$, corresponding to
a distance of roughly 1 cm out to 3 m.   
The far-left of the plot represents
the constraint coming from the search\cite{stable} for a {\em stable} $t^\prime $ (at distances
greater than approximately 3 m).

For the bounds on the $b^{\prime}$ mass, CDF assumed that the 
branching ratio $B(b^\prime \rightarrow b+Z)$ was 100 percent.
In Fig. 2, we plot the branching ratio $B(b^\prime \rightarrow b+Z)$
as a function of $m_{b^\prime}$ for different values of $m_{t^\prime}$. 
\begin{figure}
\includegraphics[angle=0,width=9cm]{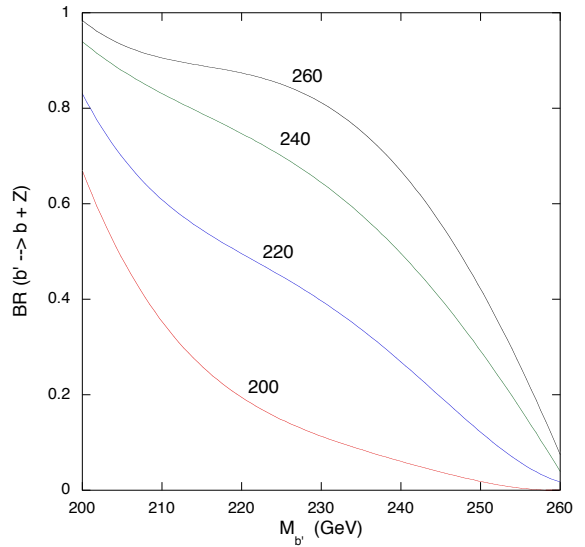}
\caption{$B(b^\prime \rightarrow b+Z)$ as a function of
$M_{b^{\prime}}$ for various $M_{t^{\prime}}$}
\end{figure}
Here we use the results of \cite{fh} where the assumption
$\sin\theta_{bt^{\prime}} = - \sin \theta_{t b^{\prime}} =x$ was made
resulting in a GIM cancellation when $m_{t^\prime} \sim m_t$. Furthermore,
as in \cite{fh}, we will assume that $|\sin \theta_{c b^{\prime}}| < x^2$
so that the decay of $b^{\prime}$ into ``lighter'' quarks will be mainly
into the $t$ quark.  Note that this assumption may not be justified, 
and if it is false the branching ratio will be even lower, weakening 
the CDF bound even further.   The decay into $t$ is three-body for
$m_{b^\prime} < m_{t} + m_{W}$ and two-body otherwise. This is to be
compared with $b^\prime \rightarrow b+Z$. This analysis has been
performed in \cite{fh} (Table I) and \cite{fhs} (Fig. 14). 

The results are shown in Fig. 2. It can be seen that 
$B(b^\prime \rightarrow b+Z) < 100 \%$
for a wide range of $b^\prime$ mass above 200 GeV. Note that
the bound of $268$ GeV  
on $m_{b^\prime}$ assuming 
$B(b^\prime \rightarrow b+Z) = 100 \%$ does not hold.
As $m_{b^\prime}$ gets above $200$ GeV, the decay mode
$b^\prime \rightarrow t + W^{*}$ begins to be comparable with
the mode $b^\prime \rightarrow b+Z$ and starts to dominate for
$m_{b^\prime} \geq 250$ GeV \cite{fh}.  In particular, for  $m_{b^\prime} > 
255$ GeV, the decay $b^\prime \rightarrow t + W$ will be into real 
particles, and thus this decay will {\em always} dominate.  The CDF 
bound can thus never exceed $255$ GeV.   Using Fig. 2, we estimate the
acceptance for $b^\prime \rightarrow b+Z$ as had been done by
CDF \cite{cdfb}. The results are then used as inputs into our analysis 
of the bounds on the $b^{\prime}$ mass.

The $b^\prime$ decay is treated in a similar fashion. Again, we subdivide
the decay into two regions: (I) $m_{b^\prime} \leq m_{t^\prime}$, and (II)
$m_{b^\prime} > m_{t^\prime}$. 
\begin{figure}
\includegraphics[angle=0,width=9cm]{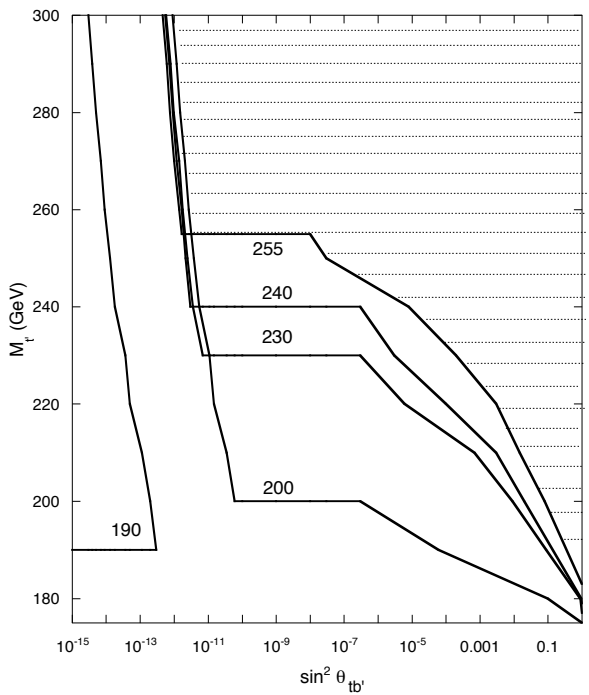}
\caption{Bound on the $b^{\prime}$ mass in the 
$m_{t^\prime}-\sin^{2}\theta_{tb^{\prime}}$ plane.}
\end{figure}
Different curves in the
$m_{t^\prime}-\sin^{2}\theta_{t b^{\prime}}$ plane correspond to
different values of $m_{b^{\prime}}$ for which $b^\prime$ {\em does not} decay
into $t^\prime$. This is shown in Fig. 3. 
Here, we take into account
the value of $B(b^\prime \rightarrow b+Z)$ (denoted by $\beta$ in
\cite{cdfb}) as obtained from Fig. 2
for a given $m_{t^\prime}$ and  $m_{b^\prime}$. We then use this number
to obtain the acceptance as given by \cite{cdfb} which is
scaled by a factor $1-(1-\beta)^2$. Different values
of $m_{b^\prime}$ for different curves in Fig. 3 reflect this
acceptance constraint. We also show an ``unexplored region''
similar to that shown in Fig. 1 for decays occuring between 1 cm
and 3 m, as well as the region where $b^\prime$ is ``stable''.
This unexplored region is not vertical, as in Fig. 1, since the rate for $b^{\prime}
\rightarrow b + Z$ is very sensitive to the $t^\prime$ mass.

In summary, we reexamined the experimental bounds on the masses of the
fourth generation quarks: the $t^\prime$ and $b^\prime$ quarks.
We divide the search into three distance regions as measured
from the center of the beam pipe: 1) $d$ = 0 cm to $\sim$ 1 cm, 
2) $d \sim$ 1 cm to 3 m and 3) $d > 3$ m. The first region
is one where most searches at the Tevatron have been performed.
We have computed the lower bounds on the $t^\prime$ and $b^\prime$
masses under the requirement that  $t^\prime$ and $b^\prime$
decay primarily into quarks of the first three generations
as shown in Fig. 1 and Fig. 3. For $b^\prime$, we found
that the CDF lower bound on its mass can never exceed
$255$ GeV, contrary to an earlier claim of $268$ GeV
which had made use of the assumption 
$B(b^\prime \rightarrow b+Z) = 100 \%$  and which
is not correct when the $b^\prime$ mass exceeds $200\,GeV$. 
For $t^\prime$, bounds are shown, starting with the CDF bound
$256$ GeV. Region (3) (greater than 3 m) is bounded by searches
for stable quarks. 
Region (2) (between 1 cm and 3 m) is {\em unexplored} and 
corresponds to a range of mixing angle $\sin^{2}\theta_{bt^{\prime}} \sim 6\times 
10^{-15}$ and $\sin^{2}\theta_{bt^{\prime}} \sim 2\times 10^{-17}$.
Such a small mixing angle might seem unlikely, but it could arise very naturally in
a 3 + 1 scenario.  For example, if one simply had a $Z_2$ symmetry in which the
fourth generation fields were odd and all other fields were even, then the mixing
angle would vanish.  However, discrete symmetries will generally be broken by
Planck mass effects, which can lead to $\sin^2\theta_{bt'}$ of $M_W/M_{Pl} \sim
10^{-17}$.   Thus, such a small mixing angle could be natural, and we urge our
experimental colleagues to explore this region.  If the fourth generation quarks
are indeed found in this region, it would shed light on the question of family
replication.

\begin{acknowledgments}
PQH is supported by the US Department
of Energy under grant No. DE-A505-89ER40518. 
MS is supported by the NSF under grant No. PHY-0554854.  We thank 
David Stuart for useful communications.
\end{acknowledgments}

\end{document}